\documentclass[aps,floatfix,twocolumn]{revtex4}

\usepackage{amsmath}
\usepackage{amssymb}
\usepackage{amsfonts}
\usepackage[dvips]{graphicx}
\usepackage{tabularx}

\begin{document}

\title{Massive phonons and gravitational dynamics in a photon-fluid model}

\author{Francesco Marino}
\affiliation{CNR-Istituto Nazionale di Ottica and INFN, Sez. di Firenze, Via Sansone 1, I-50019 Sesto Fiorentino (FI), Italy} 

\date{\today}

\begin{abstract}
We theoretically investigate the excitation dynamics in a photon-fluid with both local and nonlocal interactions. We show that the interplay between locality and an infinite-range nonlocality gives rise to a gapped Bogoliubov spectrum of elementary excitations which, at lower momenta, correspond to massive particles (phonons) with a relativistic energy-momentum relation. In this regime and in the presence of an inhomogeneous flow the density fluctuations are governed by the massive Klein-Gordon equation on the acoustic metric and thus propagate as massive scalar fields on a curved spacetime. We finally demonstrate that in the non-relativistic limit the phonon modes behave as self-gravitating quantum particles with an effective Schr\"{o}dinger-Newton dynamics, although with a finite-range gravitational interaction and a non-zero cosmological constant. Our photon-fluid represents a viable alternative to BEC models for "emergent-gravity" scenarios and offers a promising setting for analogue simulations of semiclassical gravity and quantum gravity phenomenology.
\end{abstract}


\maketitle

\section{Introduction}

Analogue gravity models provide a powerful test-bed for several aspects of classical and quantum field theories in curved spacetime \cite{rev,barcelorev,facciorev}. The general idea is that under appropriate conditions the elementary excitations in condensed-matter systems evolve as fields on a curved spacetime induced by the medium. The paradigmatic example is provided by sound waves in an inhomogeneous flowing fluid \cite{unruh,visser}. In spite of the fact that the background fluid is non-relativistic, the elementary excitations of the flow (phonons) experience a curved spacetime: their evolution is governed by the Klein-Gordon equation for a massless particle in a curved background, whose geometry is specified by a Lorentzian metric tensor (acoustic metric). As a result, the phonon dynamics exhibits an effective Lorentz invariance with the local speed of sound playing the role of the speed of light.
The coefficients of the acoustic metric depend on the fluid density, which determines also the sound speed, and the flow velocity. Hence, by tailoring the properties of the flow it is possible to simulate gravitational spacetimes and related phenomena, such as e.g. Hawking radiation, superradiance and cosmological particle production. 

Analog-gravity scenarios have been proposed and realized in a variety of physical systems, including Bose-Einstein condensates (BECs) \cite{garay2001,lahav2010}, surface waves \cite{waterwaves}, supefluid $^3$He \cite{volovik} and Fermi liquids \cite{giovanazzi}, dielectrics, \cite{reznik2000,schutzhold2002}, moving- and nonlinear-optical media \cite{leonhardt2000,philbin2008,petev} and exciton-polariton condensates \cite{gerace,nguyen}. Signatures of the Hawking process have been reported in different setups \cite{steinhauer2014,steinhauer2016,rousseaux2008,faccio2010a,faccio2011,weinfurtner2011,drori} and 
in a recent experiment the observation of superradiance has been achieved \cite{torres}.

As an alternative to the above systems, photon-fluids have recently attracted considerable attention. 
Photon fluids belong to the family of the so-called “quantum fluids of light” \cite{carusottorev}, together with exciton-polariton and photon BECs \cite{weitz}. While the last two are driven-dissipative systems based on nonlinear optical cavities, photon fluids simply rely on the nonlinear propagation of light.
A laser beam propagating through a self-defocusing medium can be described in terms of a weakly interacting Bose gas, where the repulsive photon-photon interaction arises from a third-order nonlinearity \cite{rica1992,rica1993} and the propagation coordinate acts as an effective time variable. Recent experiments in these systems provided evidence of collective many-photon phenomena, such as superfluidity and its breakdown \cite{vocke2016,michel} and nonequilibrium precondensation of classical waves \cite{santic}. In analogy with BECs, the collective excitations of the mean flow (i.e. small ripples of the transverse optical field) propagate according to the Bogoliubov dispersion relation \cite{chiao}, as recently demonstrated in thermo-optical \cite{vocke2015} and Kerr media \cite{glorieaux}. As a consequence, for the longer wavelengths a Lorentz invariant, phononic regime takes place where sound-like waves propagate with a constant speed determined by the photon fluid density. The latter is proportional to the optical intensity while the background flow velocity is controlled via the gradient of the phase profile. All these features make these systems particularly suitable for the realization of analogue gravity experiments \cite{marino2008,barad,fouxon,elazar,marino2009,braidotti,vocke2018,ornigotti,prainSR}.

All the above systems are generally characterized by a gapless dispersion relation at small momenta \cite{note1} typical of \emph{massless} collective excitations. Therefore most of the theoretical research in this area, and all ongoing experiments, have naturally focused on the simulation of massless fields propagating through a curved spacetime. A notable exception is the work by Visser and Weinfurtner \cite{silke} who first proposed a method to give rise to a spectrum of massive relativistic particles in a BEC system. The model describes a two-component BEC with an additional Raman coupling which deformes the spectrum of normal-mode excitations. Interestingly, in appropriate conditions one of the two phonon modes remains massless while the second acquires effective mass. 
Subsequent investigations by Liberati and co-workers introduced a modified BEC Hamiltonian with an $U(1)$ symmetry-breaking term \cite{eg2}. This modification provides a mass to the excitations and gives rise to a kind of analogue gravitational dynamics. Remarkably, the gravitational potential is sourced by (a function of) the density distribution of the excitations which thus play the role of the matter in this system. These are important extensions with respect to usual analogue models, since such massive excitations fields could enable simulations of quantum-gravity phenomenology (e.g. Lorentz-violating dispersion relations \cite{eg1}) and emergent-gravity scenarios \cite{eg3}. 

In the following we consider a photon-fluid model for light propagating in a defocusing medium with both local and nonlocal optical nonlinearities. At difference with the purely local case, the first-order excitations satisfy a massive version of the Bogoliubov dispersion relation in a Bose gas, with the nonlocal term being the mass-generating mechanism. For the longer wavelengths, the spectrum approximates that of a massive particle with a relativistic energy-momentum relation and, in the presence of inhomogeneous flows, the density fluctuations are described by the massive Klein-Gordon equation on the acoustic metric, thus closely mimicking the propagation of massive scalar fields on a curved spacetime.
Even more importantly, in the non-relativistic limit the phonons behaves as massive, self-gravitating quantum particles: their dynamics obeys the Schr\"{o}dinger equation in a gravitational potential whose source depends on the phononic mass density distribution via a modified Poisson equation. Unlike the Newtonian theory, we find that the range of the gravitational interaction is finite and a cosmological constant is also present. In spite of these significant differences with respect to standard gravity, such photon fluid nonetheless remains an interesting workbench for analogue simulations of semiclassical gravity scenarios. Most analogue-gravity models indeed are dealing with massless excitations that in Newtonian theory cannot act as sources of a gravitational field. This system is thus one of the very few in which a form of semiclassical gravitational dynamics can be shown to emerge.

The paper is organized as follows. In Sec. II, we introduce the modified Nonlinear Schr\"{o}dinger Equation (NSE) with local and nonlocal nonlinearities and the related photon-fluid model. We then derive the Bogoliubov-de-Gennes equations governing the dynamics of the first-order fluctuations of the optical field and the corresponding dispersion relation for a generic nonlocal function. In Sec. III we focus on a thermo-optical nonlocal nonlinearity, showing that in the defocusing case the photon fluid is stable and allows for the propagation of massive phonons, while it undergoes a Jeans instability and supports tachyonic excitations in the focusing case. The rest of the paper is devoted to analyze the fully stable defocusing regime. In Sec. IV, we address the problem of inhomogenous flows and derive a massive Klein-Gordon equation on the acustic metric that will provide the basis for the subsequent discussion on the emergent gravitational scenario. In Sec. V, we introduce the Newtonian limit of the acoustic metric which allows us to identify the gravitational potential with inhomogeneities in the photon fluid density. We then derive the non-relativistic phonon dynamics from the Klein-Gordon equation for the optical field excitations. Finally, we show that a (modified) Poisson's equation for the potential is encoded in the backreaction equation describing the first corrections to the mean-field dynamics induced by the fluctuations. The conclusions are presented in Sec. VI. 

\section{Photon fluid model and elementary excitations}

The propagation of a monochromatic optical beam oscillating at angular frequency $\omega$ in a 2D nonlinear medium can be described within the paraxial approximation in terms of the Nonlinear Schr\"{o}dinger Equation (NSE) \cite{boyd} 
\begin{equation}
\partial_z E = \frac{i}{2 k} \nabla^{2} E - i \frac{k}{n_0} E \Delta n (\vert E \vert^{2}, {\bf r}, z) \;
\label{eq1}
\end{equation}
where $E$ is the slowly varying envelope of the optical field, $z$ is the propagation coordinate, $k=2 \pi n_0 /\lambda$ is the wavenumber, $\lambda$ the vacuum wavelength and $n_0$ is the linear refractive index. The laplacian term $\nabla^{2} E$ defined with respect to the transverse coordinates ${\bf r}=(x,y)$ accounts for diffraction and $\Delta n$ is the nonlinear optical response of the medium.
For a local (Kerr) defocusing nonlinearity, $\Delta n = n_2 \vert E \vert^{2}$ with $n_2>0$, Eq. (\ref{eq1}) is formally
identical to the 2D Gross-Pitaevskii equation for a dilute boson gas with repulsive contact interactions, where the optical field $E$ corresponds to the complex order parameter and the intensity-dependent refractive index $\Delta n$ provides the interaction potential. The dynamics takes place in the transverse plane $(x,y)$ of the laser beam so that the propagation coordinate $z$ plays the role of an effective time variable $t=(n_0/c)z$, where $c$ is speed of light in vacuum.
We remark that the analogy between photon-fluids and condensates is here limited to the level of the mean-field evolution equations: as such, the system is purely classical and the optical field would correspond to the ground state wavefunction of a BEC at zero-temperature.

We consider an optical medium with both local and nonlocal third-order nonlinearities $\Delta n(\vert E \vert^{2},{\bf r},z) = n_2 \vert E \vert^{2} + \hat{\mathrm{n}}_{nl}\vert E \vert^{2}$, where $\hat{\mathrm{n}}_{nl}$ is the convolution operator 
\begin{equation}
\hat{\mathrm{n}}_{nl} \equiv \gamma \,(R \ast \,\,\,\,) = \gamma\,\theta\int d\mathbf{r}'dz' R(\mathbf{r}-\mathbf{r}',z-z')
\label{eq2}
\end{equation}
with $\gamma$ being a coefficient that depends on the specific nonlocal process and $R({\bf r},z)$ is the medium response function.
In the following we take $n_2>0$, since local repulsive interactions are required to observe a dynamically stable photon fluid on which sound waves can propagate, while $\theta = 1 (-1)$ corresponds to a defocusing (focusing) nonlocal term, respectively. 

The optical nonlocality originates from the fact that the nonlinear change in refractive index at any given position depends both on the local and on surrounding field intensity through the convolution kernel $R$. Similar nonlinear responses arise in semiconductor materials with both Kerr and thermo-optical nonlinearities \cite{torner}, nematic liquid crystals with competing orientational and thermal effects \cite{warenghem} and appear also in BECs with simultaneous local and long-range (e.g. dipolar) interactions \cite{fattori,vardi}.

The corresponding hydrodynamic formulation of the NSE is obtained by means of the Madelung transform $E = \rho^{1/2} e^{i \phi}$,
\begin{eqnarray}
\partial_t \rho + {\bf \nabla}\cdot (\rho {\bf v}) = 0 \label{eq3a}\;  \\
\partial_t \psi + \frac{1}{2} v^2 = -\frac{c^2}{n_0^3}n_2 \rho - \frac{c^2}{n_0^3} \hat{\mathrm{n}}_{nl}\rho + \frac{c^2}{2 k^2 n_0^2}\frac{\nabla^2 \rho^{1/2}}{\rho^{1/2}} \label{eq3b} 
\end{eqnarray}
where the optical intensity $\rho$ corresponds to the fluid density and ${\bf v} = \frac{c}{k n_0}\nabla \phi \equiv \nabla \psi$ is the flow velocity. On the right-hand side of (\ref{eq3b}), the first term provides the local repulsive interactions related to the positive bulk pressure $P =\frac{c^2 n_2}{2 n_0^3} \rho^2$. The second one gives rise to a nonlocal interaction potential, while the last term, directly related to diffraction, is the analogue of the Bohm quantum potential whose gradient corresponds to the so-called quantum pressure. 

We finally observe how a $z$-dependent response function would actually lead to a "non-causal" photon-fluid, as the nonlocal interactions would depend on both directions of $z=(c/n_0)t$. Such non-causality is actually fictitious since it originates from the mapping of the spatial $z$-direction into a time-coordinate. However, it suggests that a safe interpretation of the propagation coordinate in terms of a time variable would require a $z$-independent response kernel.

\subsection{Bogoliubov-de Gennes equations and excitation spectrum}

The first-order complex fluctuations $\varepsilon({\bf r},t)$ of the optical field can be described in terms of Bogoliubov excitations on top of the photon fluid. Linearizing Eq. (\ref{eq1}) around a background solution, $E = E_0(1 + \varepsilon + ...)$ with $E_0=\rho_0^{1/2} e^{i \phi_0}$, we obtain the nonlocal Bogoliubov-de Gennes equations 
\begin{eqnarray}
(\partial_T -i \frac{c}{2 k n_0} \partial_S) \varepsilon = - i \frac{\omega}{n_0} (n_2 + \hat{\mathrm{n}}_{nl})\rho_0(\varepsilon + \varepsilon^*) \label{eq4a}\;  \\
(\partial_T +i \frac{c}{2 k n_0} \partial_S) \varepsilon^* = i \frac{\omega}{n_0} (n_2 + \hat{\mathrm{n}}_{nl})\rho_0(\varepsilon + \varepsilon^*) \label{eq4b}
\end{eqnarray}
in which we have defined the usual comoving derivative $\partial_T = \partial_t + \bf{v_0} \cdot \nabla$, with ${\bf v_0}=\frac{c}{k n_0} \nabla \phi_0$, and the spatial differential operator $\partial_S = \frac{1}{\rho_0} \nabla \cdot (\rho_0 \nabla ~ ~)$.
In the spatially homogeneous case where both the background density $\rho_0$ and velocity $\bf{v_0}$ do not depend on the transverse coordinates, the plane-wave solutions of Eqs. (\ref{eq4a})-(\ref{eq4b}) satisfy the dispersion relation
\begin{equation}
\Omega^2 = c_s^2 K^2(1 + \theta \frac{\gamma}{n_2}\tilde{R}(K,n_0\Omega/c) + \frac{\xi^2}{4\pi^2}K^2).
	\label{bogo}
\end{equation}
where $K$ is the wavenumber of the mode, $\Omega = \Omega' - {\bf K}\cdot{\bf v_0}$ its angular frequency in the locally-comoving background frame and $\tilde{R}$ is the three-dimensional Fourier transform of the response function $R({\bf r},z)$.  We remark that here the angular frequency $\Omega'$ actually corresponds to the longitudinal wavenumber $K_z$ expressed in temporal units via $\Omega'=K_z c/n_0$ while $\bf{K}$=($K_x$,$K_y$) is the transverse wavevector. 

In analogy to purely local BECs and photon-fluids \cite{marino2008}, we defined in Eq. (\ref{bogo}) the sound speed $c_s^2 \equiv \frac{d P(\rho_0)}{d \rho} = \frac{c^2 n_2}{n_0^3} \rho_0$ and the healing length, $\xi=\lambda/2\sqrt{n_0 n_2 \rho_0}$ as the characteristic length separating the linear (phononic) and quadratic (single-particle) regime of the dispersion relation for $\gamma=0$.

The length $\xi$ determines the critical wavenumber $K_c=2\pi/\xi$ associated to the breakdown of Lorentz invariance, generally expected to occur in quantum gravity phenomenology at the Planck scale. Low energy modes with $K\ll K_c$ propagate indeed at the invariant universal speed $c_s$, while at higher momenta $K \gg K_c$ the terms arising from the quantum pressure become dominant and the group velocity of excitations increases with $K$.

Within the paraxial approximation and in the presence of nonlocal processes with negligible longitudinal dependence, the main contribution to the non-locality comes from the $K$-dependence of $\tilde{R}$ and we can thus safely assume $\tilde{R}(K,n_0\Omega/c)\simeq \tilde{R}(K,0)$. This is indeed the case of thermo-optical nonlinearities dominated by the transverse diffusion of heat \cite{vocke2015,danieleSN} that we will discuss in the next sections. From now on we shall ignore the z-dependence of $R$.

We finally observe that on the basis of Eq. (\ref{bogo}) wave propagation for focusing nonlinearities $\theta=-1$ is allowed only for wavenumbers $K$ such that $\Omega^2>0$, i.e. $\tilde{R}(K)< \frac{n_2}{\gamma}(1+K^2 / K_c^2)$. Negative values of $\Omega^2$ corresponds to exponentially-growing modes characteristic of linearly-unstable flows.
In the defocusing case $\theta=1$, the system is neutrally stable to perturbations of all wavenumbers, hence supporting travelling waves. While the plane wave solution is always modulationally stable, instabilities and wave-breaking phenomena \cite{conti2} are expected in the presence of inhomogeneous beams and/or discontinuous response kernels. In spite of this fact, stable operation in nonlocal photon fluids has been experimentally demonstrated even in the presence of background inhomogeneities \cite{vocke2016,vocke2018}.

\section{Thermo-optical nonlocality and massive excitations}

The functional form of $\tilde{R}(K)$ depends on the specific nonlocal process under consideration. A case of particular interest is provided by light propagation in thermo-optical media, where the change of refractive index $\hat{\mathrm{n}}_{nl} \rho =n_{th}$ arises from the temperature increase due to the residual laser absorption. The heat diffuses through the material and eventually across the boundaries of the medium. As a result, the shape of the response function will strongly depend also on the transverse boundary conditions \cite{Minovich2007}. This might open interesting perspectives in experiments since it could be possible to tailor the nonlocal response of the medium, e.g. by acting on the geometry of sample, in order to modify the dispersion (\ref{bogo}) and in turn the physical properties of the collective excitations \cite{vocke2016}.

In the limit of an infinite medium in the two transverse dimensions $n_{th}$ is coupled to the optical intensity through the stationary heat equation \cite{Swartz,carmon}
\begin{equation}
- \nabla^{2} n_{th} = \frac{\alpha \vert \beta \vert}{\kappa} \rho 
\label{eq5}
\end{equation}
where $\kappa$ is the thermal conductivity of the material, $\alpha$ its linear absorption coefficient and $\beta = \vert \partial n_{th} / \partial T \vert$ is the change in the refractive index with respect to the temperature. 
The heat equation (\ref{eq5}) dictates that the corresponding range of the nonlocal interactions between photons is infinite (infinite-range nonlocality) \cite{carmon}.
The nature of the nonlinearity (focusing or defocusing, leading to attractive or repulsive interactions) depends on the sign of $\beta$. Here we take the absolute value $\vert \beta \vert$ since the sign is already considered in (\ref{eq2}) by the coefficient $\theta$. 

Fourier transforming the expression $n_{th}=\gamma \theta R({\bf r)} \ast \rho$ and Eq. (\ref{eq5}) one can readily verify that $\tilde{n}_{th} = \gamma \theta \tilde{R}(K)\tilde{\rho}=\frac{\alpha \vert \beta \vert}{\kappa K^2} \tilde{\rho}$ and thus $\tilde{R}(K) \propto 1/K^2$. This implies that the convolution integral $\hat{\mathrm{n}}_{nl} \rho$ is, up to a constant, the solution of Eq. (\ref{eq5}) with $R({\bf r})$ being the Green's function of the 2D Laplacian operator. Hence, the following relation holds: $\gamma \theta \nabla^2 R({\bf r}) =-(\alpha \vert \beta \vert/\kappa) \delta({\bf r} - {\bf r'})$.

\subsection{Defocusing nonlocal nonlinearity: massive phonons}

Using the above $\tilde{R}(K) = \frac{\alpha \vert \beta \vert}{\kappa \gamma K^2}$ in Eq. (\ref{bogo}) and considering a defocusing nonlinearity $\theta=1$ we find
\begin{equation}
\Omega^2 = \Omega_0^2 + c_s^2 K^2\left(1 + \frac{\xi^2}{4\pi^2}K^2\right)
\label{disp1} 
\end{equation}
where $\Omega_0=c \sqrt{\frac{\alpha \vert \beta \vert}{\kappa n_0^3} \rho_0}$ has indeed the dimensions of a frequency. 
Hence, we can thus identify $\hbar \Omega_0$ with the rest energy of a particle and write $\hbar \Omega_0 = m c_s^2$, where $m$ is the rest mass and $c_s$ plays the role of the light speed. Defining the excitation momentum $p=\hbar K$ and the critical momentum $p_c=\hbar K_c= h/\xi$, Eq. (\ref{disp1}) can be rewritten in the form
\begin{equation}
\mathcal{E}^2 = m^2 c_s^4 + c_s^2 p^2\left(1 + \frac{p^2}{p_c^2}\right) \,\,.
\label{disp2}
\end{equation}
The properties of the above dispersion strongly depend on the thermo-optical coefficients of the material used to produce the photon fluid. Here we are interested in investigating the regime in which $p_c \gg m c_s$.
In this case, Eq. (\ref{disp2}) is a generalization of the Bogoliubov dispersion relation describing \emph{massive} collective excitations with high-energy, Lorentz-violating corrections. Similar modified dispersion laws with extra momentum-dependent terms appear in several phenomenological approaches to quantum gravity, where $p_c$ is typically associated to the Planck momentum \cite{slv}.

The dispersion curve (\ref{disp2}) interpolates between three different regimes depending on the fluctuations momentum.

When $p \gg p_c$, the quartic term dominates and Eq. (\ref{disp2}) approximates the free-particle behaviour $\mathcal{E} \approx c_s p^2/p_c$. Using the above definitions of $p_c$ and $c_s$ and the photon momentum $p_{\gamma}=\hbar n_0 k$ we get $\mathcal{E} \approx c p^2/(2p_{\gamma})$ (or equivalently $\mathcal{E} \approx p^2/2 m_{\gamma}$ introducing an effective photon mass $m_{\gamma}=p_{\gamma}/c$). Therefore in analogy to BEC analogue models, the excitation energy tends to the energy of the individual particles forming the background fluid, i.e. in our case the photons.

In the intermediate regime, $m c_s \lesssim p \ll p_c$, we obtain the "relativistic" dispersion relation for a massive particle $\mathcal{E} \approx \sqrt{p^2 c_s^2 + m^2 c_s^4}$, with the speed of sound playing the role of the speed of light. These are collective excitations exactly like usual phonons in local quantum fluids, but possessing a finite rest mass.

At lower momenta, $p \ll m c_s$, the phonon modes enter the non-relativistic regime: the energy-momentum relation reduces to $\mathcal{E} \approx p^2/2m + m c_s^2$, where the first term is the kinetic energy of a particle of mass $m$ and the second
its constant rest mass energy.

We notice that the rest frequency $\Omega_0$ depends only on the strength of the thermo-optical (nonlocal) nonlinearity. The latter is thus responsible for the generation of the gap in the dispersion relation and hence for the onset of the excitation mass. On the other hand, the "invariant" limit speed $c_s$ is determined solely on the local defocusing effect. 

The rest frequency $\Omega_0$ can also be expressed in terms of the sound speed as $\Omega_0=c_s \sqrt{\frac{\alpha \vert \beta \vert}{\kappa n_2}}=c_s / \lambdabar_C$. The characterisitic length-scale $\lambdabar_C$, given by the square root of the ratio between the local and nonlocal coefficients, corresponds to the acoustic analogue of the reduced Compton wavelength of the particle $\lambdabar_C=\hbar/(m c_s)$. The inverse of this length defines the above non-relativistic limit through $K \ll \lambdabar_C^{-1}$ and, as we shall see in Sect. V, it provides also some of the fundamental characteristic scales of the emergent gravitational force.
 
We conclude the section briefly discussing a more realistic model of the thermo-optical nonlinearity \cite{vocke2016,conti1,conti2} given by 
\begin{equation}
-\nabla^{2} n_{th} + \frac{n_{th}}{\sigma^2} = \frac{\alpha \vert \beta \vert}{\kappa} \rho 
\label{dls}
\end{equation} 
in which the effects of the distant boundaries has been included in the distributed loss term $n_{th} / \sigma^2$, where $\sigma$ is the length-scale of the nonlocal interaction. 

Eq. (\ref{dls}) allows us to continuosly describe the transition from an infinite-range to finite-range thermo-optical nonlocality and has provided a theoretical framework for the phenomenological Lorentzian response adopted in previous experiments (see e.g. Refs. \cite{vocke2015,barad}). The Fourier-transformed response associated to (\ref{dls}) $\gamma \tilde{R} = \frac{\alpha \vert \beta \vert}{\kappa}\frac{\sigma^2}{1 + \sigma^2 K^2}$ has indeed a Lorentzian shape, where $2/\sigma$ is its full-width at half-maximum. The dispersion (\ref{disp2}) is thus modified as
\begin{equation}
\mathcal{E}^2 = m^2 c_s^4 \frac{p^2}{p_{nl}^2 + p^2} + c_s^2 p^2\left(1 + \frac{p^2}{p_c^2}\right)
\label{disp3}
\end{equation}
where we introduced the nonlocal momentum $p_{nl}=\hbar/\sigma$. The above response kernel reduces to the ideal form of the infinite space model $\gamma \tilde{R}(K) = (\frac{\alpha \vert \beta \vert}{\kappa}) / K^2$ in the limit of $\sigma K \gg 1$. Such a regime can be reasonably reproduced by means of suitable background optical beams comprising wavevectors only of $K \gg 1/\sigma$. This procedure has been implemented in a lead-doped glass experiment \cite{danieleSN}. Before being launched into the nonlinear medium, the laser beam has been passed through a phase mask generating a ring-shaped beam with zero intensity at $K=0$ and large-enough transverse wavevectors. Using this technique, the authors demostrated a nonlocal thermo-optical nonlinearity with $\sigma K \approx 20$.
In this case $p \gg p_{nl}$, and Eq. (\ref{disp3}) well approximates the massive Bogoliubov dispersion (\ref{disp2}). 
However, for finite $p_{nl}$ the two relations will eventually differ at arbitrarily low momenta, as the gap in (\ref{disp3}) arises only in the singular limit of an infinite-range nonlocality, $p_{nl}=0$ ($\sigma \rightarrow \infty$).

\subsection{Focusing nonlocal nonlinearity: Jeans instability}

For $\theta=-1$ the hydrodynamic equations Eq. (\ref{eq3a})-(\ref{eq3b}) together with Eq. (\ref{dls}) describe a ($2+1$)-dimensional quantum fluid with local repulsive and finite-range attractive interactions. In the ideal case of an infinite medium the model reproduces the nonlinear evolution of a self-gravitating BEC \cite{jones}, where the nonlocal change of refractive index $n_{th}$, solution of Eq. (\ref{eq5}), mimics a Newtonian potential generated by the fluid mass density. In the absence of local interactions, Eq. (\ref{eq1})-(\ref{eq5}) are indeed formally equivalent to the Schr\"{o}dinger-Newton (SN) equation in two spatial dimensions \cite{segevSN,danieleSN}, originally proposed by Diosi \cite{diosi} and Penrose \cite{penrose} as a model for quantum wave function collapse (see also \cite{bassiSN} for further discussion). 

Concerning the dynamics of elementary excitations, in the more general case the dispersion relation reads
\begin{equation}
\mathcal{E}^2 = -m^2 c_s^4 \frac{p^2}{p_{nl}^2 + p^2} + c_s^2 p^2\left(1 + \frac{p^2}{p_c^2}\right) \,\,.
\label{disp4}
\end{equation}
The negative sign in front of the rest-energy term originated from to the attractive photon interactions gives rise to two fundamentally different behaviours at high and low momenta. The critical wavenumber $K=K_{J}$ separating these two regimes is implicitly defined by the condition at which the wave frequency (energy) vanishes,
\begin{equation}
\frac{\sigma^2}{\lambdabar_C^2} \frac{1}{1 + \sigma^2 K^2} - \left(1 + \frac{K^2}{K_c^2}\right) = 0 \,\,.
\label{cond1}
\end{equation}
For high wavenumbers $K > K_{J}$, the local repulsive interactions and the quantum pressure are sufficiently strong to countebalance the nonlocal attractive forces and the waves are freely oscillating. In the opposite case, we have growing excitation modes revealing the linear instability of the system (see also Refs. \cite{chavanis,ghosh} for a discussion in self-gravitating BECs).

In the hydrodynamic (Thomas-Fermi) approximation ($K \ll K_c$) and for $(\sigma K) \gg 1$, Eq. (\ref{cond1}) simply
reduces to the ordinary Jeans instability condition, where $K_J$ corresponds to the Compton wavenumber of the particle $K_J=\lambdabar_C^{-1}$. In astrophysics such instability is thought to be responsible for the collapse of interstellar gas clouds eventually leading to star formation. 

In the stable regime, $K > \lambdabar_C^{-1}$, and for $K \ll K_c$ Eq. (\ref{disp4}) yields the tachyonic dispersion relation $\mathcal{E} \approx \sqrt{p^2 c_s^2 - m^2 c_s^4}$, with real energy and momentum and imaginary rest mass. Excitations of any wavenumuber in fact propagate at supersonic group velocities, with the invariant $c_s$ being now a lower limit for propagation speeds. From now on we will focus on the fully stable case of a photon-fluid with defocusing non-local nonlinearity.

\section{Inhomogenous flows and massive Klein-Gordon equation}

For purely local interactions $\gamma=0$, a formal equivalence can be established between phonons propagating on top of the photon fluid and the evolution of scalar fields in curved spacetime \cite{marino2008}. The equation of motion is typically derived by linearizing Eqs. (\ref{eq3a})-(\ref{eq3b}) around a background state since phonons, i.e. the acoustic elementary excitations, are defined as the \emph{first-order fluctuations} of the quantities describing the mean fluid flow: $\rho = \rho_0 + \epsilon \rho_1 + O(\epsilon^2)$, $\psi = \psi_0 + \epsilon \psi_1 + O(\epsilon^2)$.
When the terms arising from quantum pressure are negligible, the phonon dynamics is fully described by a single second-order equation for the linearized velocity-potential which has the form of the Klein-Gordon equation for a massless scalar field
\begin{equation}
\Box \psi_1 \equiv {1\over\sqrt{-g}} \partial_\mu \left( \sqrt{-g}\; g^{\mu\nu} \; \partial_\nu\ \psi_1 \right)
\label{eq6}
\end{equation}
propagating in a (2+1)-dimensional curved spacetime whose geometry is described by the acoustic metric $g_{\mu\nu}$, with inverse $g^{\mu\nu}$ and determinant $g$,
\begin{equation}
g_{\mu\nu} =
\left(\frac{\rho_0}{c_s} \right)^2 \left(\begin{array}{cc}
  -(c_s^2 - v_0^2)  &  -{\bf v_0^T} \\
  -{\bf v_0}  &  {\bf I} \\
\end{array}
\right)
\label{metric}
\end{equation}
where ${\bf I}$ is the two-dimensional identity matrix.

The above scenario is deeply modified in the presence of both local and nonlocal nonlinearities. In this context, it is convenient to derive the acoustic metric directly from the nonlocal Bogoliubov-de Gennes equations (\ref{eq4a})-(\ref{eq4b}).

To this end, we apply the operator $(\partial_T +i \frac{c}{2 k n_0} \partial_S) (\frac{1}{\rho_0} ~ ~)$ to Eq. (\ref{eq4a}) and we obtain
\begin{eqnarray}
(\partial_T +i \frac{c}{2 k n_0} \partial_S)\frac{1}{\rho_0}(\partial_T -i \frac{c}{2 k n_0} \partial_S) \varepsilon = \frac{c_s^2}{\rho_0} \partial_S \varepsilon  \nonumber\\ 
- i \gamma (\frac{\omega}{n_0}\partial_T +i \frac{c^2}{2 n_0^3} \partial_S)\frac{1}{\rho_0}  R \ast [\rho_0(\varepsilon + \varepsilon^*)]
\label{eq7}
\end{eqnarray}
As in the local case, we remind that the gravitational analogy holds in the phononic regime in which the dispersion relation
takes the relativistic form with a limit propagation speed, i.e. for wavenumbers $K \ll K_c$. The corresponding equation for the excitation field can thus be obtained by ignoring the higher-order spatial derivatives in Eq. (\ref{eq7}), which indeed are responsible for the Lorentz-breaking, $K^4$-terms in the dispersion relation.
A close inspection of Eq. (\ref{eq7}) suggests that such approximation corresponds to take the diffractionless limit $k \rightarrow \infty$ \cite{prainSR} or, equivalently, neglect the quantum pressure terms arising from the linearized hydrodynamic equations \cite{marino2008}. In this limit Eqs. (\ref{eq3a})-(\ref{eq3b}) indeed reduce to the Navier-Stokes equations for a barotropic, irrotational and inviscid fluid, in which the Lorentz simmetry associated to phonon dynamics is not explicitly broken \cite{visser}.

Under this approximation and using the fact that the background density $\rho_0$ satisfies the continuity equation (\ref{eq3a}) with ${\bf v}$=${\bf v_0}$, Eq. (\ref{eq7}) can be re-written as 
\begin{eqnarray}
\Box \varepsilon \,= \,- i \gamma \frac{\omega}{n_0} (\partial_T+\nabla\cdot {\bf v_0})  R \ast [\rho_0(\varepsilon + \varepsilon^*)] \nonumber\\ 
+ \gamma \frac{c^2}{2 n_0^3}\nabla \cdot (\nabla - \nabla \ln \rho_0)  R \ast [\rho_0(\varepsilon + \varepsilon^*)] \
\label{eq8}
\end{eqnarray}
where 
\begin{equation}
\Box \equiv (\partial_T+\nabla\cdot {\bf v_0})\partial_T-\nabla \cdot(c_s^2\nabla \,\,\,)
\label{eq8b}
\end{equation}
is precisely the d'Alambertian operator associated with the acoustic metric $g_{\mu\nu}$.

In the purely local case $\gamma=0$, we recover the usual Klein-Gordon equation for a massless particle on curved spacetime, here described by the complex field $\varepsilon$.  For a spatially homogeneous background, with constant density $\rho_0$ and constant flow velocity ${\bf v_0}=\frac{c}{k n_0} \nabla \phi_0$, the operator $\partial_T = \partial_t + \bf{v_0} \cdot \nabla$ commutes with the convolution operation and using Eqs. (\ref{eq4a})-(\ref{eq4b}) we find that the wave equation (\ref{eq8}) becomes independent of $\varepsilon^*$, 
\begin{equation}
(\partial_{TT}^2- c_s^{2} \nabla^2)\varepsilon  = c_s^2 \frac{\gamma}{n_2}R(r) \ast \nabla^2 \varepsilon \,.
\label{eq9}
\end{equation}
It is immediate to verify that the Fourier transform of Eq. (\ref{eq9}) leads to the dispersion law (\ref{bogo}).

The complex fluctuations $\varepsilon$ can be easily linked to the real density and phase perturbations though the relations $\rho_1=\rho_0(\varepsilon+\varepsilon^*)$ and $\phi_1=(i/2)(\varepsilon^*-\varepsilon)$. By means of these expressions and using the relation between the optical phase and velocity-potential of the flow, $\psi_1= (c/k n_0) \phi_1$, one can split Eq. (\ref{eq8}) into the following system of wave-equations 
\begin{eqnarray}
\Box \psi_1 = -\gamma \frac{c^2}{n_0^3} (\partial_T + \nabla \cdot {\bf v_0}) R \ast \rho_1 \label{eq10a}\;  \\
\Box \left(\frac{\rho_1}{\rho_0}\right) = \gamma \frac{c^2}{n_0^3} \nabla \cdot (\nabla - \nabla \ln \rho_0) R \ast \rho_1 \
\label{eq10b}
\end{eqnarray}
For local fluids $\gamma=0$ Eq. (\ref{eq10a}) reduces to the massless Klein-Gordon equation (\ref{eq6}) for the velocity-potential perturbations $\psi_1$ and an equation of the same form is satisfied also by the relative density fluctuations $\rho_1/\rho_0$. 

In the ideal case of infinite-range thermo-optical nonlinearity, $\sigma \rightarrow \infty$, the response function satisfies $\gamma \nabla^2 R({\bf r}) =-(\alpha \vert \beta \vert/\kappa) \delta({\bf r} - {\bf r'})$. Using this result and considering a nearly homogeneous background density \cite{note2}, Eq. (\ref{eq10b}) takes the form of the massive Klein-Gordon equation in curved spacetime
\begin{equation}
\Box \rho_1 + \Omega_0^2 \rho_1 = 0 \,.
\label{eq11}\;  \\
\end{equation}
In the more realistic case of finite-range thermo-optical nonlinearities, Eq. (\ref{eq11}) remains basically valid for perturbations with wavenumbers $1/\sigma \ll K \ll K_c$. 

\section{Emergent gravitational dynamics}

In the previous sections we have seen that phonon excitations in our system behave as massive particles with a relativistic energy-momentum relation. In the presence of inhomogeneous flows we also derived a massive Klein–Gordon equation on the acoustic metric for the density fluctuations that thus reproduce the evolution of massive scalar fields on a curved spacetime. The spacetime curvature which mimics the gravitational field arises from the inhomogeneity of the background, whose dynamics however is governed by a non-relativistic nonlinear equations (cf Eq. (\ref{eq1}) or, equivalently, Eqs. (\ref{eq3a})-(\ref{eq3b})). As such the analogy works only at the \emph{kinematical} level: the fluctuations propagate in a given background solution associated to a specific spacetime configuration. All effects due to gravitational back-reaction are neglected, i.e. the spacetime geometry is not modified by the perturbations propagating on it. While under certain conditions it is possible to extend the analogy and include in a geometric framework even the evolution of the background \cite{goulart1,goulart2,marino2016}, there is no possibility in general to describe the \emph{dynamics} of the acoustic metric in terms of something similar to Einstein's equations. The situation changes if one considers relativistic models and interesting progresses in this direction have been made, e.g. in the framework of relativistic BECs \cite{eg4}. 

Nevertheless, as mentioned before a kind of gravitational dynamics may emerge even in non-relativistic BECs upon suitable
modifications of the standard equations to break the $U(1)$ symmetry associated with the conservation of particle number \cite{eg2}. 
In such a modified model, the massive excitations feel a Newtonian gravitational potential whose source is related to the excitation density. 

In the following we show that a similar scenario arises also in our nonlocal photon-fluid. 

\subsection{The Newtonian limit}

In Sect. IV we treated the general case of an inhomogeneous background, i.e. of an arbitrary curved spacetime simulating a generic gravitational field. 
Since the background dynamics is non-relativistic we expect to find at most a kind of Newtonian gravity, as previously shown in other non-relativistic frameworks \cite{eg2}.
We thus focus on a nearly-homogeneous background corresponding to a weak gravitational field because, in analogy to the weak-field approaximation of General Relativity (GR), is in this limit that a Newtonian-like gravity is expected to emerge.

In GR the weakness of gravitational field allows for the decomposition of the metric into a flat Minkowski spacetime, $\eta_{\mu\nu}$, plus a small perturbation $\emph{g}_{\mu \nu}=\eta_{\mu \nu}+h_{\mu\nu}$, $\vert h_{\mu\nu}\vert \ll 1$ where $h_{\mu\nu}$ represents the weak deviations from flatness $g_{\mu \nu}$ i.e. the gravitational field. In this regime the Newtonian potential is related to the metric though the equation $g_{00}=\eta_{00}+ h_{00}\simeq -(1 + 2\Phi_N/c^2)$, which follows from a non-relativistic limit of the geodesics equation \cite{wald}.

In analogy to the above, we consider a photon fluid with zero-flow and a spatially-localized inhomogeneity in the density, i.e. $E_0=\rho_{\infty}^{1/2}(1 + u(r))$ with $u \ll 1$ and  $u\rightarrow 0$ at infinity. We thus assume that only a small region of the fluid deviates from the constant asymptotic value of the density $\rho_{\infty}$. 
This implies a rescaling of the speed of sound $c_s^2=c_{\infty}^2(1 + 2u(r))$ and thus of the $00$-component of the acoustic metric (\ref{metric}). On the basis of the hydrodynamic equations (\ref{eq3a})-(\ref{eq3b}), a density inhomogeneity would also imply an inhomogeneous flow. However as demonstrated in Ref. \cite{eg2}, the velocity perturbations do not contribute \emph{at first order} to the anologue gravitational potential. In other words, at the first order all the information about the gravitational potential is encoded in the density perturbation. This result is general in acoustic models and does not depend on the specific fluid under consideration. Therefore for the sake of simplicity and without loss of generality, here we assume deviations in the density only. 

\subsection{Non-relativistic phonon dynamics}

In order to show the emergence of a gravitational potential term we should derive the equation of motion for excitations in the non-relativistic regime of the Bogoliubov spectrum, i.e. when $p \ll m c_{\infty}$, and for an homogeneous background except for the small density-inhomogeneity $u(r)$.

To this end, we start directly from the nonlocal wave-equation (\ref{eq8}) for the complex excitation field $\varepsilon$. Setting $\rho_0=\rho_{\infty}(1 + 2u(r))$ and ${\bf v_0}=0$ we get
\begin{equation}
\Box \varepsilon  = c_{\infty}^2 \frac{\gamma}{n_2}\nabla^2 R(r) \ast [(1 + 2u(r))\varepsilon]
\label{eq12}
\end{equation}
In deriving (\ref{eq12}) we disregarded terms containing the spatial derivatives of $u(r)$ and the products $u(r)\nabla^2 \varepsilon$. The former are negligible in the asymptotic region and the latter are suppressed both by the smallness of $u$ and by the fact that we are interested into the non-relativistic regime $p \ll m c_{\infty}$.
Restricting ourselves to the case of an infinite-range thermo-optical nonlinearity, Eq. (\ref{eq12}) further simplifies and reads
\begin{equation}
(\partial_{tt}^2 - c_{\infty}^2 \nabla^2)\varepsilon + \Omega_{\infty}^2(1 + 2u(r))\varepsilon = 0
\label{eq13}
\end{equation}
where we have defined the asymptotic rest frequency $\Omega_{\infty}=c_{\infty} \sqrt{\alpha \vert \beta \vert/\kappa n_2}$. 

The non-relativistic limit $p \ll m c_{\infty}$ (or equivalently, $c_{\infty} \rightarrow \infty$) means that the kinetic energy of the particle should be small with respect to its mass energy $\hbar \Omega_{\infty}=mc_{\infty}^2$. Making the ansatz $\varepsilon=\varphi exp(-i\Omega_{\infty}t)$ to factor out the rest frequency (i.e. the contribution to the total energy due to the rest energy of the particle) we can approximate \cite{greiner}
\begin{equation}
\partial_{tt}^2 \varepsilon \simeq (-2i\Omega_{\infty} \partial_t \varphi- \Omega_{\infty}^2 \varphi)
e^{-i \Omega_{\infty} t} \nonumber\\ \,\, .
\end{equation}
Substituting the above expression into Eq. (\ref{eq13}) we get the Schr\"{o}dinger equation for a particle of mass $m$
\begin{equation}
i \hbar \partial_{t} \varphi = -\frac{\hbar^2}{2 m}\nabla^2 \varphi + mc_{\infty}^2u(r)\varphi
\label{eq14}
\end{equation}
subject to an external potential proportional to $u(r)$. The latter can be formally identified as a gravitational potential defining $\Phi_{G}=c_{\infty}^2u(r)$. We finally remark that for finite-range thermo-optical nonlinearities, Eq. (\ref{eq14}) would remain approximately valid in the momentum range $p_{nl} \ll p \ll m c_{\infty}$.

\subsection{Modified Poisson equation}

In the previous section we have found that the low-energy evolution of massive phonons obeys the Schr\"{o}dinger equation for a nonrelativistic quantum particle in an external potential $\Phi_{G}$. Our interpretation of $\Phi_{G}$ as a gravitational potential is based on the way it enters in the Schr\"{o}dinger equation and because it is related to the $00$-component of the acoustic metric, similarly to the Newtonian potential in the weak-field approximation of GR. 
In this framework we should find that in the appropriate limits $\Phi_{G}$ also obeys a kind of Poisson's equation \cite{note3}.

Following the same argument of Ref. \cite{eg2}, since the Newtonian potential is the manifestation of
small deviations of the order parameter $E$ from perfect homogeneity, the corresponding Laplace's equation should be encoded in the nonlinear evolution equation (\ref{eq1}). On the other hand, we expect the source term to be directly related to the phononic fluctuations. Indeed in Newtonian gravity the only source for the gravitational field is a mass-density distribution and in our system this can originate only from the massive elementary excitations.
As a result, the non-relativistic massive phonons should experience a kind of gravitational potential generated by  themselves, i.e. they should feel their own gravity. 

This self-interaction can thus be derived from the nonlinear equation Eq. (\ref{eq1}), but adding the corrections to the mean-field dynamics induced by the fluctuations, in order to see how the phonons backreact over the background fluid. 

Backreaction effects 
can be calculated expanding Eq. (\ref{eq1}) up to second order, $E = E_0 + \eta_1 + \eta_2 =E_{B} + \eta_1$: here $\eta_1$ and $\eta_2$ are linear and quadratic quantities in the fluctuation amplitude, respectively, and we introduced new variable $E_{B}$ including the modifications to the zero-order dynamics, that is to say the backreaction \cite{balbinot}. 

Substituting the above ansatz in Eq. (\ref{eq1}) with $\Delta n = n_2 \vert E \vert^{2} + \gamma R \ast \vert E \vert^{2}$ we obtain
\begin{eqnarray}
\partial_t(E_{B}+\eta_1)=\frac{i c}{2k n_0} \nabla^{2}(E_{B}+\eta_1)-\frac{i \omega}{n_0}(E_{B}+\eta_1)\Delta n_B  \; \nonumber\\
\label{eq15}
\end{eqnarray}
where $\Delta n_B = (n_2 + \gamma R \ast )(\vert E_{B}\vert^2 + 2 \mathrm{Re}(E_{B}^{*}\eta_1) + \vert \eta_1 \vert^{2})$ and we used the effective time coordinate $t=(n_0/c)z$.

Since $E_{B}$ consists only of zero-order and second-order terms in the fluctuation amplitude, all linear quantities in $\eta_1$ must vanish. The zeroing of the linear fluctuation terms in (\ref{eq15}) leads to the Bogoliubov-de Gennes equations (\ref{eq4a})-(\ref{eq4b}) upon substituting $\eta_1=E_0\varepsilon$ and $E_{B} = E_0$. 
What remains is a nonlinear evolution equation for $E_{B}$ in which the fluctuations appear quadratically
\begin{eqnarray}
\partial_t E_{B}&=&\frac{ic}{2k n_0} \nabla^{2}E_{B} - \frac{i\omega}{n_0}E_B[(n_2 + \gamma R \ast )\vert E_{B}\vert^2  + 2 n_2\vert\eta_1\vert^{2}] \nonumber \\ &-& \frac{i\omega}{n_0}[n_2 E_{B}^* \eta_1^2 + \gamma (E_{B} R\ast\vert\eta_1\vert^2 + 2\eta_1 R \ast \mathrm{Re}(E_B \eta_1^*))] \; \nonumber\\
\label{eq16}
\end{eqnarray}
Here, $\eta_1$ is in general time-dependent and for the purpose of simulating the effects of quantum and/or thermal fluctuations on the mean-field dynamics in analogy to real quantum gases, it could be taken as a stochastic variable. 
However, for our purposes we are now interested in calculating the backreaction effects on stationary background solutions, i.e. stationary spacetime geometries. The corresponding mean-field equation is then obtained by time-averaging Eq. (\ref{eq16}) and replacing the fluctuating quantities with their mean values. For $\gamma=0$, the averaged (\ref{eq16}) closely resembles the modified Gross-Pitaevskii equation with beyond-mean-field corrections due to fluctuations \cite{giorgini1,giorgini2,salasnich}. In this context the quadratic terms $\mathfrak{n}({\bf r})=\langle\vert \eta_1 \vert^{2}\rangle$, and $\mathfrak{m}({\bf r})=\langle \eta_1^2\rangle$ play the role of the density of non-condensed particles (i.e. the "out-of-condensate" photons) and of the anomalous density in the Bogoliubov-Popov-Beliaev approximation. The quantities $\gamma \langle \eta_1 R \ast \mathrm{Re}(E_B^{*} \eta_1) \rangle$ and $\gamma R \ast \langle\vert \eta_1 \vert^{2}\rangle$ provide further corrections due to nonlocality. 


Setting $\langle E_B \rangle = \rho_{\infty}^{1/2}(1 + u(r))$ with $u \ll 1$ in the averaged Eq. (\ref{eq16}) yields 
\begin{eqnarray}
\frac{c}{2k n_0}\nabla^{2} u - \frac{\omega}{n_0}\rho_{\infty}(n_2 + \gamma R \ast )(1+2u)= \nonumber\\ 
\frac{\omega}{n_0}[n_2(2 \mathfrak{n}({\bf r}) + \mathfrak{m}({\bf r})) + \gamma(R \ast \mathfrak{n}({\bf r}) + 2 \mathfrak{g}({\bf r}))]
\label{eq17}
\end{eqnarray}
where $\mathfrak{g}({\bf r})=\langle \eta_1 [R \ast \mathrm{Re}(\eta_1^*)]\rangle$. 

In the ideal case of our interest of infinite-range thermo-optical nonlinearity, and applying the operator $c/(2 k n_0)\nabla^2$ to both sides of Eq. (\ref{eq17}), we get
\begin{eqnarray}
\frac{1}{K_c^2}\nabla^{4} \Phi_G - \nabla^{2} \Phi_G + \frac{1}{\lambdabar_C^2}\Phi_G + \frac{c_{\infty}^2}{2 \lambdabar_C^2} = \nonumber\\
 \frac{c_{\infty}^2}{2 \lambdabar_C^2 \rho_{\infty}} [\lambdabar_C^2 \nabla^2 (2\mathfrak{n}({\bf r}) + \mathfrak{m}({\bf r})) - \mathfrak{n}({\bf r}) + 2\nabla^2 \mathfrak{g}({\bf r})]  
\label{eq18}
\end{eqnarray}
where here in the definition of the critical wavevector $K_c= 2 \pi/\xi_{\infty}$ we used the asymptotic healing length 
$\xi_{\infty}=\lambda/2\sqrt{n_0 n_2 \rho_{\infty}}$. 

Eq. (\ref{eq18}) can be interpreted as a modified fourth-order Poisson equation for the potential $\Phi_G$, provided that we are able to identify the right-hand side with a genuine source term. To this end, we remark that $\mathfrak{n}$, $\mathfrak{m}$ and $\mathfrak{g}$ are quadratic terms in $\eta_1$, i.e. they have the dimension of a photon-fluid mass-density and are related to the phononic excitations through the relation $\eta_1=E_0\varepsilon$. Therefore, they can be safely interpreted as a source of the gravitational field. 

In contrast to Newtonian gravity, Eq. (\ref{eq18}) contains the additional terms $\nabla^{4}\Phi_G/K_c^2$ and $\Phi_G/\lambdabar_C^2$. Owing to the $K_c^2$ coefficient, the first term suggests modifications of the Poisson equation which would become increasingly important as the analogue of the Planck scale $\xi_{\infty}$ is approached. Since in the nonrelativistic limit here considered we are dealing with long-wavelength modes with $K \ll K_c$, the variations of $\Phi_G$ over spatial scales of order $\xi_{\infty}$ can be neglected. 

The second term $\Phi_G/\lambdabar_C^2$ denotes a finite range for the gravitational interaction, with a characteristic length scale given by $\lambdabar_C$. Such a finite interaction scale for gravity would translate into a massive graviton with a mass that, in our model, corresponds to that of the massive phonons. A further comparison with the Newtonian limit of Einstein equations allows us to identify the quantity $1/2 \lambdabar_C^2$, which indeed has the dimension of the square of an inverse length, with a cosmological constant $\Lambda$. 
Finally, in the right hand side of the Eq. (\ref{eq18}) it is natural to define the analogous of the universal gravitational constant $G=\frac{c_{\infty}^2}{2 \lambdabar_C^2 \rho_{\infty}}$. 

In light of the above considerations, Eq. (\ref{eq18}) takes the more meaningful form 
\begin{equation}
\nabla^{2} \Phi_G - \frac{1}{\lambdabar_C^2}\Phi_G = c_{\infty}^2\Lambda + G \varrho_{matter}({\bf r})
\label{eq19}
\end{equation}
where we introduced the mass-density distribution
\begin{equation}
\varrho_{matter}({\bf r})= -\lambdabar_C^2 \nabla^2 [2\mathfrak{n}({\bf r}) + \mathfrak{m}({\bf r})] + \mathfrak{n}({\bf r}) - 2 \nabla^2 \mathfrak{g}({\bf r}) 
\label{eq20}
\end{equation}
The mass-density (\ref{eq20}) is a complicated function of quadratic fluctuation-terms, which deserves further analysis. We first note that in the non-relativistic limit, $\lambdabar_C^2 K^2 \ll 1$, the first laplacian-term is less relevant with respect to the others and can thus be neglected. Moreover, applying the product rule to $\nabla^2 \mathfrak{g}({\bf r})$ we find
\begin{eqnarray}
\nabla^2 \mathfrak{g}({\bf r}) &=& \langle [R \ast \mathrm{Re}(\eta_1^*)] \nabla^2\eta_1 \rangle + 2 \langle \nabla \eta_1 \cdot \nabla [R \ast \mathrm{Re}(\eta_1^*)]\rangle \nonumber\\ 
&-& \frac{1}{2}[\mathfrak{m}({\bf r}) + \mathfrak{n}({\bf r})]
\label{eq21}
\end{eqnarray}
where we used the previously defined quantities $\mathfrak{n}({\bf r})=\langle\vert \eta_1 \vert^{2}\rangle$, and $\mathfrak{m}({\bf r})=\langle \eta_1^2\rangle$ and the relation $\nabla^2 R({\bf r}) = -\delta({\bf r} - {\bf r'})$, valid in the limit of infinite-range nonlocality. While the first two terms in (\ref{eq21})are difficult to handle in general, we observe that their contribution becomes less important at very low wavevectors, in which case the mass-density distribution approximates 
\begin{equation}
\varrho_{matter}({\bf r})\approx 2\mathfrak{n}({\bf r}) + \mathfrak{m}({\bf r}) 
\label{eq22}
\end{equation}
The above expression coincides with the density distribution obtained in Ref. \cite{eg2} (see Eq. 43) where a Newtonian-like gravity has been shown to emerge in a BEC model modified with a $U(1)$ symmetry-breaking term. The distribution (\ref{eq22}) has an immediate physical interpretation: in analogy to BECs, the two terms $\mathfrak{n}({\bf r})$ and $\mathfrak{m}({\bf r})$ indeed correspond to the so-called normal and anomalous density encoding the effects of "non-condensed particles" on the mean-field dynamics \cite{giorgini1,giorgini2,salasnich}. 

We finally remark that, similarly to the rest energy, also the Newton constant can be expressed solely in terms of the nonlocal coefficients, $G=(c^2/2 n_0^3)(\alpha \vert \beta \vert/\kappa)$. The nonlocal nonlinearity is thus responsible for the emergent gravitational interaction.

To conclude, the Eqs. (\ref{eq14}) and (\ref{eq19}), mutually coupled via the relations $\eta_1=E_0\varepsilon$ and (\ref{eq22}), actually gives rise to an effective Schr\"{o}dinger-Newton dynamics describing the evolution of a quantum mass density experiencing its own gravitational field \cite{bassiSN}, although here the source is a more complicated function of the mass-density distribution and the gravitational interaction is characterized by a short interaction-range and a nonzero cosmological constant. 

\section{Conclusions and future perspectives}

Quantum fluids of light such as exciton-polaritons BECs and more recently photon fluids have offered alternative platforms for fundamental studies of quantum many-body physics. Recent experiments in these systems provided evidence of collective many-photon phenomena, such as the emergence of a phonon-regime in the Bogoliubov dispersion \cite{vocke2015,glorieaux}, superfluidity and nucleation of quantized vortices in the flow past a physical obstacle \cite{vocke2016,michel} and classical wave condensation \cite{santic}.

Here, we have theoretically investigated a photon-fluid with both local and nonlocal interactions from the analogue gravity perspective. We have found that collective excitations in this system display a gapped Bogoliubov spectrum which at low energies correspond to that of massive phonons with a relativistic energy-momentum relation. In the presence of an inhomogeneous flow the dynamics of the density fluctuations is equivalent to that of a massive scalar field propagating in a curved spacetime whose geometry is specified by the acoustic metric. This generalizes previous studies in local fluids to the case of massive phonons and provides a quite natural setting for analogue simulations of quantum gravity phenomenology.

The massive nature of the elementary excitations allows us to study their nonrelativistic dynamics in nearly-homogeneous background that, as explained, corresponds to the case of a weak gravitational field. In this limits we find that the phonon-modes behave as a \emph{self-gravitating} quantum system. The evolution equations are indeed the Schr\"{o}dinger equation for a massive quantum particle, including a term that represents the interaction of the particle with its own gravitational field, and a kind of Poisson equation with a source depending on the phononic mass-density distribution. In analogy to the Newtonian limit of GR, the potential in the Poisson equation is related to the background geometry (namely to the 00-component of the metric) experienced by the particles propagating on it.
Since most of analogue models are dealing with massless excitations that in the framework of Newtonian gravity cannot act as sources of a gravitational field, our system is one of the very few in which a form of semiclassical gravitational dynamics can emerge. 

One of the next stages of this investigation will focus on the design of realistic experimental schemes for the implementation of such photon-fluids. Apart from the analogue-gravity side, we expect these experiments to be interesting also from the perspective of the quantum fluids of light, as the interplay between local and nonlocal nonlinearities with different --possibly tunable-- kernels could unveil new collective many-photon phenomena and hydrodynamic phase transitions.

\section*{Acknowledgements}

I wish to thank Daniele Faccio and Antonello Ortolan for stimulating discussions and useful comments on the manuscript.


\begin{thebibliography}{99}
\bibitem{note1} The confinement of photons into a cavity such as in polariton BECs gives rise to gapped dispersion relations although with imaginary terms due to the intrinsically dissipative nature of the cavity configuration (see e.g.\cite{gerace,carusottorev}. Here we are referring to (formally) conservative analog gravity models.
\bibitem{note2} On the basis of the fluid equations (\ref{eq3a})-(\ref{eq3b}), a nearly-constant density would also imply a nearly-homogeneous flow. Nevertheless, there are specific situations in which the simplifying assumption to consider a nearly-constant density while letting the flow vary provides an useful model of the system (e.g. in the case of vortex flows when the locations of the event horizons and the ergosphere occur in regions of nearly constant density, far away from the vortex core).
\bibitem{note3} We remark that the weak-field limit of GR in ($2+1$)-dimensions (i.e. the dimensionality of our system) does not correspond to 2D Newtonian gravity. In two space dimensions all vacuum solutions of Einstein's equations are flat and even in the presence of matter, the spacetime curvature remains null everywhere except in the interior of the matter distribution. Similarly to other analogue gravity models, such a restiction does not apply to our system whose spacetime metrics are not solutions of the Einstein field equations. As a consequence, the Newtonian-like dynamics here discussed should thus be hypothetically considered as the limit of a theory that strongly differs from Einstein's gravity.
\end{thebibliography}
\end{document}